\documentclass[aps,prd,twocolumn,groupedaddress]{revtex4}
\usepackage{graphicx}
\usepackage{amssymb,amsmath,amsfonts,palatino,amsthm}
\usepackage{amssymb}
\usepackage{epstopdf}
\newcommand{\f}{\begin{equation}}
\newcommand{\ff}{\end{equation}}

\newcommand{\beq}{\begin{equation}}
\newcommand{\eeq}{\end{equation}}

\begin{document}


\title{Could quantum mechanics be an approximation to another theory?}


\author{Lee Smolin}
\email[lsmolin@perimeterinstitute.ca]{Your e-mail address}
\affiliation{Perimeter Institute for Theoretical Physics,\\
Waterloo, Ontario N2L 2Y5, Canada,\\ and \\
Department of Physics, University of Waterloo,\\
Waterloo, Ontario N2L 3G1, Canada}


\date{\today}

\begin{abstract}

We consider the hypothesis that quantum mechanics is an approximation to another, cosmological theory, accurate only for the description of subsystems of the universe.  Quantum theory is then to be derived  from the cosmological theory by averaging over variables which are not internal to the subsystem, which may be considered non-local hidden variables.  We find conditions for arriving at quantum mechanics through such a procedure. The key lesson is that the effect of the coupling to the external degrees of freedom introduces noise into the evolution of the system degrees of freedom, while preserving a notion of averaged conserved energy and time reversal invariance.  

These conditions imply that the effective description of the subsystem is  Nelson's stochastic formulation of quantum theory. We show that Nelson's formulation is not, by itself,  a  classical stochastic theory as the conserved averaged energy is not a linear function of the probability density.   We also investigate an argument of Wallstrom posed against the equivalence of Nelson's stochastic mechanics and quantum mechanics and show that, at least for a simple case,  it is in error.

\end{abstract}


\maketitle
\tableofcontents
\section{Introduction}

In spite of much progress clarifying foundational issues in quantum mechanics, there remains persistent evidence  that quantum mechanics is an approximation to a deeper theory.   Among the reasons for this belief are;

\begin{itemize}

\item{}The unresolved difficulties in extending quantum theory to cosmology.  If this cannot be done then one possible explanation is that quantum mechanics does not in fact extend to the whole universe. It must then be an approximation to a more fundamental cosmological theory, which applies only for small subsystems of the universe.

\item{} The  difficulties in solving the measurement problem in the context of a theory with a realistic ontology.

\item{} The success of quantum information theory, which reinforces the viewpoint that the quantum state represents the information that observers have of a system.

\item{} The experimental evidence against the Bell inequalities tells us that any theory quantum mechanics is derived from must be non-local. It is then natural to hypothesize that this non-local theory is a cosmological theory, which is more adequate than  quantum mechanics for the investigation of cosmological problems.  

\end{itemize}

These assertions raise  issues about which there is disagreement and lively discussion.  However, apart from taking the fact that these issues are unresolved as motivation,  the  aim of this paper is not to address that debate directly.  Instead, we ask whether there are conditions that must be satisfied by any non-local, cosmological theory, so that quantum mechanics can be derived from it as an approximate description of subsystems. 

Given that locality is sometimes thought to be an almost sacred principle, it  is perhaps an under-appreciated fact that there are non-local hidden variables theories which reproduce the predictions of quantum mechanics. The best studied of these is Bohmian mechanics\cite{Bohm}. Others more recently proposed include the trace  dynamics of Adler\cite{Adler} and the approach described in \cite{hidden}.  

In Bohm's theory the ontology includes both the particle positions and the wavefunction, both of which live in the classical configuration space.  If one believes that the particles are real one must also believe the wavefunction is real because it determines the actual trajectories of the particles. This allows us to have a realist interpretation which solves the measurement problem, but the cost is to believe in a double ontology.  

Is it possible to solve the measurement problem with a realistic ontology that  is not doubled, as Bohm's is?  The idea that quantum mechanics is an approximation to a
non-local, cosmological theory offers a new possibility for doing this because 
the missing information, which makes quantum theory statistical, would be found, not in a more detailed description of a subsystem, as it is in Bohmian mechanics, but in hidden variables which describe relationships between the subsystem and the rest of the universe. 

Such a theory would begin with a non-quantum theory giving a precise description of a whole universe. The dynamics of this cosmological theory might be deterministic or stochastic, but it must be non-local.  One would choose a subsystem and then represent the rest of the universe, including possible non-local variables relating the subsystem to the rest of the universe, by a statistical ensemble.  Averaging over this ensemble one would hope to arrive at quantum mechanics as a description of the subsystem.  Formulations of such a theory have been studied in \cite{hidden,morehidden} and  appear to work, at least to a leading approximation.   

In fact, there is  independent motivation for believing that non-local interactions arise from microscopic theories of spacetime  These arise from the observation that locality can disordered in quantum states of geometry in
loop quantum gravity and similar theories\cite{fotini-daniele,me-daniele,Hal}.  This is possible because in such theories the classical metric is an emergent, approximate quantity, meaningful only at low energies.  As a result, there can be a mismatch between the notion of locality which controls the dynamics of the fundamental microscopic theory and a different notion of locality that follows from the emergent metric which controls the approximate, low energy, dynamics.  This can happen in both quantum and stochastic dynamics of discrete geometries\cite{Hal}. Given the latter possibility it is possible to conjecture that a stochastic theory of pre-geometry is the cosmological theory which quantum theory approximates for subsystems.

In this paper we ask what conditions must be imposed so that a derivation along the lines just sketched would work.  We begin with a classical Hamiltonian description of the cosmological theory. This is probably not necessary, but we employ it for definiteness.  The subsystem phase space variables will be given by the canonical pairs, $(x^a,p_b)$ while
the remaining variables, both those local to the universe outside the subsystem and non-local variables connecting them, are denoted by pairs $(y^\alpha , \pi_\beta )$. 
We will assume that the universe as a whole  can be described statistically by means of  a  joint probability distribution
$\tilde{\rho}(x,y; p,\pi,  t)$,  which evolves according to some kind of non-quantum evolution
which could be classically deterministic or stochastic.   

We then want to know what conditions must be imposed on the total system dynamics, or on the interaction between the subsystem and hidden variables, such that quantum mechanics can be derived as an approximate description of the subsystem.  

To investigate this it is natural to  employ the tools of stochastic differential equations as an intermediate step between the classical statistical mechanics of the cosmological theory and the quantum mechanical description  of the subsystem.  This allows us to make use of  the formulation of quantum theory in the language of stochastic dynamics, formulated by Nelson\cite{Nelson-PhysRev} and studied by he\cite{Nelson-books} and others\cite{Nelson-others}.   Nelson's formulation is employed, not as a fundamental reformulation of quantum theory, but as a step in a derivation of quantum theory from a hidden variables theory. In the language of effective field theory we can say that Nelson's stochastic formulation of quantum mechanics is a kind of low energy, approximate description of a small subsystem of the universe. 

Having described the method, I can state the conclusions, which are a list of conditions required for quantum mechanics to be derived from a non-local cosmological theory as a description of a small subsystem of the universe.    

\begin{itemize}
 
\item{}{\bf P1} Averaging over the hidden variables induces a random noise in the evolution of the subsystem variables. This allows the subsystem to be described in terms of stochastic differential equations.    

\item{}{\bf P2} There is a notion of averaged energy in the subsystem, which is conserved in time.

\item{}{\bf P3} The averaging over the effects of the hidden variables does not impose a direction of time. After the hidden variables are averaged over the stochastic dynamics of the subsystem is 
 invariant under time reversal. 

\item{}{\bf P4 } We can approximate the averaged dynamics by keeping terms only of lowest order in spacial and time derivatives.  In particular, the probability current for the subsystem variables is assumed to be irrotational.  

\end{itemize}

The fourth  condition is natural if we take the point of view in which we are extracting an approximation good for low energies.  From this point of view it is natural to ask that the probability current be irrotational, because in many cases of interest, such as fluid flow, this condition can be satisfied by lowering the energy.

We note that the second and third of these are very stringent conditions.  It is often the case that  when a subsystem is coupled to a reservoir, the subsystem degrees of freedom become stochastic. But usually the dynamical laws satisfied by  their averaged quantities are neither conservative nor time reversal invariant.  Our conditions say that quantum mechanics could be an approximate description of a subsystem of a non-local theory only if these unlikely conditions 
are nevertheless satisfied.  

This paper is organized as follows. We begin in the next section by presenting a formulation of Nelson's stochastic mechanics suitable for the purposes at hand.  We do this to set the stage and also because employing Nelson's formulation in this way requires that we confront and resolve some questions that have been raised about its adequacy as a formulation of quantum dynamics.  We do so in the next section and in section IV. 

In the next section we clarify the use  of probability in Nelson's formulation. In classical or quantum statistical mechanics the average of any observable is linear in the probability density.  We show that this is not the case in Nelson's formulation, because the averaged energy, which is conserved, depends non-linearly 
on the probability density in configuration space.  More specifically, the energy of a system of particles at a point in their configuration space depends on the gradient of the probability density at that point.  This makes Nelson's formulation similar to Bohm, in that  the probability density, which is a property of an ensemble, is playing a role influencing the motion of individual systems. 

If there is to be a clean derivation of quantum theory from a non-local hidden variables theory it must explain  this non-linear dependence on the probability density  of subsystem variables in terms of a linear  dependence on the probability density of the non-local hidden variables. The key question is whether and when this is possible.  This is the task of section III where we will provide conditions under which the formulation of section II may be derived as an approximation to a more fundamental theory. We see that the key condition that must be satisfied for this to be the case  is time reversal invariance.

Finally, in section IV we discuss a criticism of Nelson's formulation given by 
Wallstrom\cite{Wallstrom} and show that, in the simple case of quantum mechanics on a circle, it leads to the wrong conclusion.

\section{Quantum theory as a form of stochastic dynamics}

 The basic idea of Nelson is that quantum physics
arises from an irreducible Brownian motion on the configuration space of a system which
has, however, the unusual property that it is {\it non-dissipative.}  That is, in spite of the fact that there is an irreducible noise in the evolution of the configuration variables, there is still an
average conserved energy.   In this section we want to understand clearly how this  is
possible.  We employ an effective  Hamiltonian approach as this allows us to understand
Nelson's formulation as a low energy effective description of a more fundamental theory. 

\subsection{The implementation of stochastic dynamics.}

We  start with a review of the basic formulation of stochastic 
differential equations. For more details see \cite{stochastic,Nelson-PhysRev,Nelson-books}.

We consider a system consisting of degrees of freedom, $(x^a)$
where $a=1,...N$,  We assume there is a metric $q_{ab}$ on the configuration space
of the $x^a$'s.  We posit a  probability 
distribution $\rho (x^a , t)$ and a probability current $v^a (x,t)$,
related by the conservation law
\f
\frac{d \rho}{dt} = - {\partial \rho v^a \over \partial x^a } 
\label{conserve}
\ff
This preserves the normalization condition that probabilities add up to one.
\f
\int d^nx \rho (x )   =1
\label{isone}
\ff

We assume also that the evolution of the $x^a$'s is subject to a 
noise, so that it must be described by a stochastic differential 
equation,
\beq
dx^a=b^a dt +dw^a,
\eeq
with 
\beq
\left\langle dw^a dw^b \right \rangle=2\nu dt q^{ab}.
\eeq
for $dt>0$, 
with diffusion constant $\nu$.

For $dt<0$, we have instead, 
\beq
dx^{*a}=b^{*a} dt +dw^{*a},
\eeq
with
\beq
\left\langle dw^{*a} dw^{*b} \right \rangle=-2\nu dt q^{ab}.
\eeq

These imply the Fokker-Planck equations
\begin{eqnarray}
\dot{\rho} &= & -\nabla_a (\rho b^a ) + \nu \nabla^2 \rho \nonumber \\
\dot{\rho} &= & \nabla_a (\rho b^{*a} ) - \nu \nabla^2 \rho
\end{eqnarray}

If we define the current velocity $v^a$ by 
\beq
v^a= {1\over 2} (b^a + b^{*a} ) 
\eeq
we recover the conservation law for probability (\ref{conserve}). 

We then define the difference of the forward and backwards velocities
to be  the osmotic velocity
\beq
u^a= {1\over 2} (b^a -  b^{*a} )  
\label{udef}
\eeq
The Fokker-Planck equations then imply
\f
u^a =  \nu q^{ab} \nabla_b \ln \rho
\ff
This extremely interesting relation will turn out to be the key to the mystery of how stochastic mechanics works. It tells us that the difference between the forward and backwards current velocity is related to the derivative of the probability density. It is important to note that this is a general property of Brownian motion.  It also reminds us that, like $\rho$, $b^a$ and $b^{*a}$ are properties, not of individual histories of the system, but of an ensemble of histories. 

We are used to the idea that in statistical mechanics averages over  ensembles give us predictions for where we are likely to find an individual,typical system in the ensemble. But we see that in Brownian motion something a bit more is involved, because properties of the probability distribution at one time give us information about the dynamics of the ensemble and hence, to some extent about the future or past behavior of individual systems. 

We will be later interested in the properties of these observables under time reversal invariance $t \rightarrow -t$
\beq
b^a \rightarrow -b^{*a} , \ \ \ \ v^a \rightarrow -v^a , \\ 
u^a \rightarrow u^a
\label{timereverse}
\eeq

These last equations, (\ref{conserve}) to (\ref{timereverse}) define a general 
system subject to a brownian motion and so realizes {\bf P1}.

\subsection{The implementation of the average conserved energy condition}

We now
describe how we implement the condition {\bf P2-P4}. The first of these asserts that  there is  an 
equilibrium state in which there is
an averaged conserved 
energy. We require that there is a 
function $H$ of the variables that describe the stochastic
process that satisfies the following conditions.

\begin{itemize}
    
    \item{} $H$ is a function of  $\rho$,  $ v^a (x)$ and position.

    \item{} $H$ is conserved in time,  ({\bf P2}).
    
    \item{}$H$ is invariant under time reversal invariance,  ({\bf P3}). 
    
        \item{}$H$ contains only those terms that dominate in the low 
    velocity and long long wavelength  limit ({\bf P4}).
    
    \item{}The current velocity $v^a (x)$ is irrotational   ({\bf P4}).
    
    \end{itemize}
    
    It is also reasonable to impose the following condition, to ensure stability
    
    \begin{itemize}
    
       \item{} $H$ is positive definite.
    
    \end{itemize}
    
    FInally, it is natural to demand the following two conditions:
    
    \begin{itemize}

    \item{} $H$ is invariant under rotations among the $x^a$'s.
    
    \item{} $H$ is local, so it is of the form
    \f
    H= \int d^Nx \tilde{\cal H}
    \ff
    where $\tilde{\cal H}$ is a density.

\end{itemize}    

We now proceed to define the form of the effective hamiltonian from these conditions.

The only density available to us is $\rho$. Hence we have
\f
\tilde{\cal H}= \rho h
\ff
here $h$ is a scalar function of  density weight zero, of $\rho$ and $v^a$.
As $\rho$ is a density the only way $h$ can depend on
$\rho$ is by depending on $u^a$, which by (\ref{udef}) is
a vector.  As $h$ must be rotationally invariant it can 
depend only on $v^2$, $u^2$ and $v \cdot u$.  Of these $v^2$ and
$u^2$ are time reversal invariant, but $u\cdot v$ is not. Hence the lowest order 
dependence possible on $u \cdot v$ is through $(u \cdot v)^2$.  

Furthermore, $\rho$ and $v^a$ may be independently specified as
initial data, subject only to (\ref{isone}), as $\dot{\rho}$ is determined
by the conservation law (\ref{conserve}).  For the long wavelength limit we should also include terms with as few derivatives as possible. 
Hence, putting all the conditions
but the last together, we have
\begin{eqnarray}
h(x) &=&  F_1 [v^2 ] + F_2 [u^2 ] + F_3 [ (u\cdot v)^2 ]   + {\cal U}(x) 
\nonumber \\ 
&& + \mbox{ terms in }  \ \partial v \mbox{ and } \ \  \partial u
\end{eqnarray}
where the $F_i$ are positive definite functions and 
${\cal U}(x)$ is a positive, time independent,
function of the $x^a$'s, which of 
course is a potential energy. 

For small velocities we may expand $F_1$ as
\f
F_1 [v^2 ] =   {m\over 2} [ v^2 + {v^4 \over c^2} + ... ]
\ff
where $m \geq 0 $ has dimensions of mass, $c$ has dimensions of velocity, and 
$F_2$ has the same behavior. $F_3$ starts at order $(u \cdot v)^2 /c^2$.

Hence, for small velocities $h$ is then required by the condition {\bf P4} to be of the form,
\f
h(x) = {m\over 2}  v^2 + {b\over 2} u^2 + {\cal U}(x)
\label{hquad}
\ff
where $m,b \geq 0$.

We note that if $b=0$ $H$ is just the averaged energy of an ensemble 
of classical systems. Thus we can identify  $m$ as the mass. 

If we want to get something different from classical physics, we then want
$b > 0$ which gives us a term in the averaged energy
\f
H_{quantum} = \int d^Nx \rho \frac{b}{2} u^2 =  
\frac{b\nu^2}{2} \int d^Nx \rho  (\nabla \ln \rho )^2
\label{hquantum}
\ff
This term is significant because, as we will see, it gives rise to the quantity that in  Bohm's formulation is called the ``quantum potential".  It is  thus the key to how stochastic dynamics becomes quantum mechanics, and we will seek to understand it better once we have completed the derivation.

We now impose the condition that the 
motion is irrotational, so that
\f
{\partial v^a \over \partial x^b} - {\partial v^b \over \partial 
x^a}=0
\ff

This implies there is a function $S(x)$ such that
\f
v^a = {1\over m} q^{ab} {\partial S \over \partial x^b }  
\label{Sdef}
\ff

The $m$ is put there for later convenience. 

\subsection{Derivation of the Schroedinger equation}

We are now ready to give the derivation of the Schroedinger equation.

Using (\ref{udef}) the conserved average energy is now given by
\f
H= \int \rho \left ( {1\over 2m} (\nabla_a S)^2 + 
{b \nu^2 \over 2} (\nabla_a \ln \rho )^2 + {\cal U}
\right ) 
\ff
We now impose conservation of the averaged energy
\begin{eqnarray}
0= \dot{H} & = &  \int \left \{ 
\dot{\rho} [ {1\over 2m} (\nabla_a S)^2 + {\cal U} - 
{b \nu^2 \over 2} ({ ( \nabla_a \rho )^2 \over \rho^2 }  \right. \nonumber \\
&& \left. + 2\nabla^a {1\over \rho} \nabla_a \rho   )
]  - \dot{S} {1\over 2m} \nabla^a ( \rho \nabla S )
\right \} 
\end{eqnarray}
Now we use (\ref{conserve}) and (\ref{Sdef}) which 
imply, 
\f
\dot{\rho}= -{1\over m}\nabla^a (\rho \nabla_a S),
\label{conserve2}
\ff
which give us
\f
0= \int 
\dot{\rho} \left [\dot{S} + {1\over 2m} (\nabla_a S)^2 + {\cal U} - 
{b \nu^2 \over 2} ({ ( \nabla_a \rho )^2 \over \rho^2 } 
+ 2\nabla^a {1\over \rho} \nabla_a \rho   )
\right ]
\label{bohm}
\ff
This is solved for all $\dot{\rho}$'s if the integrand is zero:
\f
0=  \dot{S} + {1\over 2m} (\nabla_a S)^2 + {\cal U} - 
{b \nu^2 \over 2} ({ ( \nabla_a \rho )^2 \over \rho^2 } 
+ 2\nabla^a {1\over \rho} \nabla_a \rho   )
\label{Seq}
\ff

But this last eq (\ref{Seq}) and (\ref{conserve2}) are the real and 
imaginary parts of the Schrodinger equation. To see this, we define
$\hbar$ by 
\f
 \hbar \equiv \nu \sqrt{mb}
\label{hbardef}
\ff
We then define the wavefunction,
\f
\Psi (x,t) = \sqrt{\rho} e^{\imath S / \hbar}
\ff
It is then straightforward that eq (\ref{Seq}) and (\ref{conserve2})
are the real and imaginary parts of 
\f
\imath \hbar { d\Psi \over dt} = \left [
-{\hbar^2 \over 2m} \nabla^2 + {\cal U}
\right ]\Psi
\ff

This provides a derivation of Nelson's stochastic formulation of 
quantum theory, showing the importance of the assumptions of a positive
averaged squared energy and time reversal invariance. 

\subsection{The scene of the crime}

We see from this derivation that the key point which separates quantum mechanics from a statistical ensemble of classical, conserved systems is the term
$H_{quantum}$, eq. (\ref{hquantum}) in the averaged conserved energy.

This gives rise to  the term proportional to $b$, and hence $\hbar^2$, in (\ref{Seq}).  
In fact this is the only place $\hbar$ appears in the coupled  equations (\ref{Seq})
and (\ref{conserve2}).  
This term  is
called, in Bohm's formulation, the quantum potential.  In Bohm's theory, this is the origin
of the effect by which the wavefunction guides the motion of the particle (it is hence the pilot in the pilot-wave theory.)   The claim that Nelson's formulation provides an alternative to Bohm that is realist but without a dualist ontology rests on the claim that this term arises from the stochastic fluctuations of the particle. Let us then examine how this term appears.

In the derivation just given, 
we opened the door to the possibility of quantum theory when we posited
that the energy at a point in configuration space, $h(x)$ could be a function 
not only of position but of properties of the ensemble such as $v^a$ and $u^a$.
Is this permissible?  At first sight, it would seem an innocuous assumption.  If we want to get the dynamics of a classical ensemble of conserved systems, we have to include
in $h(x)$ a term proportional to $v^2$.   But once we have this term, why not
include also a term proportional to $u^2$?

The difference is that $v^a$ is at a single time an independent quantity, whereas
$u^a$ is determined by the Fokker-Planck equation to be equal to
$\nu \nabla^a \ln \rho $.  Thus, including terms in $u^a$ in the energy 
make it a function itself of the probability density. To avoid this we should have
set $b=0$ which would have implied $\hbar =0$.  

The issue is that when we write the definition of an ensemble average of energy
as 
\f
H= \int d^N x \rho (x) h (x)
\ff
we {\it expect} that $h(x)$ is a property of an individual trajectory, i.e. the energy of a particle at $x$, and that the properties of the ensemble are reflected completely in the
linear explicit dependence on $\rho$. This expectation would seem to rule out
$h$ being itself a function of $\rho$. However it is, if it is a function of $u^a$ and the
ensemble is Brownian.  

This makes clear that Nelson's is not simply a theory of an ensemble of particles undergoing Brownian motion.  For if, as this derivation suggests, it is meaningful to interpret 
$h(x)$ as the energy of a quantum particle at a point $x$, the energy of that particle 
depends on the gradient of the probability density at that point.  To put this another way, it costs a lot of energy for a particle to be at a point where the gradient of the probability density-in the ensemble of which it is a member- is large.  

One could take two points of view about this.  One could, first of all,  {\it simple assume that the energy of a particle can depend on the gradient of the probability} for an ensemble it is a part of.  Then we are giving the ensemble a physical, causative role it cannot have in ordinary classical physics.  This implies that Nelson's stochastic mechanics is  much closer to Bohm than seems at first, in that the particles in a system are not the only entities that must be posited to have a deterministic dynamics, for properties of the ensemble play a role in the dynamics of individual systems.  This seems to come close to giving the ensemble probabiltiy an ontological status close to, if not identical to, that of the wavefunction in Bohm's theory.  

We might on the other hand, try to argue that this non-linear dependence of the energy of a particle on $\nabla \ln \rho (x) $ results from averaging over a distribution of non-local hidden variables. This would be to explain Nelson from a deeper theory, as investigated in \cite{hidden,morehidden}. We see here that the key point of such a derivation must be to cause the dynamics of a single observable to depend non-linearly on its own probability distribution.  It is possible that this could come from averaging over an ensemble of hidden variables, as argued in \cite{hidden,morehidden}, but if so we should see in detail how it works.

This is the task of the next section. 

\section{Deriving Nelson from a hidden variables theory}

Having clarified the assumptions behind Nelson's stochastic formulation of quantum mechanics we turn to the main question of the paper,which is when and how those assumptions can be consequences of a non-local hidden variables theory. 

\subsection{Schema for a non-local hidden variables theory}

Here is one schema for how a derivation of quantum mechanics from a more fundamental theory can go: The more fundamental theory has configuration observables
$(x^a, y^\alpha)$ where $x^a$ label the configuration space of the subsystem which is to be described by quantum mechanics and $y^\alpha$ stands for the hidden variables. Conjugate to
them are momentum observables $(p_a, \pi_\alpha )$.   

The evolution of the system is governed by a joint Hamiltonian 
\f
H= H^{subsystem}(x,p) + H^{hidden} (y,\pi ) + H^{int} (x,y,p,\pi)
\ff
To evade Bell's theorem we have to assume that $H^{int}$ introduces interactions which are non-local in the configuration space of the subsystem, and we will assume this is the case.  We assume a standard form for
\f
H= H^{subsystem}(x,p)= \frac{1}{2m} q^{ab}  p_a p_b + {\cal U} (x)
\ff
where $q^{ab}$ is the inverse metric on the configuration space of the subsystem. 

The statistical state of the system is described by a probability density on
phase space $\Gamma$, $\tilde{\rho} [x,y;p,\pi ]$.  The expectation values of observables
$O[x,y;p,\pi ]$ are given by
\f
<O> = \int_\Gamma \tilde{\rho} O
\ff
To make the connection with quantum theory we will assume that the solution of the joint dynamics is given by a solution of the Hamilton-Jacobi function $S(x,y)$ on the joint
configuration space. Then we may consider a statistical distribution holding $S$ fixed
\f
\tilde{\rho}_S [x,y;p,\pi ] = \tilde{\rho}(x,y)\prod_a \delta (p_a - \frac{\delta S}{\delta x^a})
\prod_\alpha \delta (p_\alpha - \frac{\delta S}{\delta y^\alpha})
\ff
We can then write, on solutions,
\f
p_a =  \frac{\delta S}{\delta x^a}= m q_{ab}\dot{x}^b (x,y)
\ff
Where we have written $\dot{x}^b (x,y)$ to indicate that the velocities given by the solution $S(x,y)$ are functions of both the position in the configuration space of the subsystem, $x^a$ and the hidden variables $y^\alpha$.  It is this dependence that
can carry the non-local couplings necessary for a non-local hidden variables theory.

This lets us write
\begin{eqnarray}
< H^{subsystem}> &=& \int dx dy \tilde{\rho}(x,y) \left [
\frac{m}{2} q_{ab} \dot{x}^a (x,y)\dot{x}^b (x,y)  \right. \nonumber \\
&& \left.  + V(x)
\right ]
\end{eqnarray}
We will also need the probability distribution on the configuration space of the subsystem
\f
\rho (x) = \int dy \tilde{\rho}(x,y)
\ff

\subsection{Derivation of stochastic dynamics}

We can now turn our attention to the conditions needed to reproduce the derivation of quantum theory in the previous section.   We are particularly concerned with the derivation of the term that becomes the quantum potential (\ref{hquantum}).  

We will assume that 
\f
< H^{int} > =0
\ff
and that $< H^{hidden} >$ is separately conserved or may be neglected. This gives 
us a separate conservation of $< H^{subsystem}>$, i.e.
\f
\frac{d< H^{subsystem}>}{dt} =0
\ff

To investigate the implications of this we turn our attention to the  kinetic energy term
\f
k(x,y) = \frac{m}{2} q_{ab} \dot{x}^a (x,y)\dot{x}^b (x,y)
\ff
 The expectation
value of the kinetic energy is then
\f
<k> = \int dy \int d x  \tilde{\rho} (x,y) \frac{m}{2} \dot{x}^2 (x,y)
\label{trade1}
\ff

With the hidden variables specified the velocities can be assumed to exist, so that we can write
\f
k(x,y) =\frac{m}{2} \lim_{dt \rightarrow 0 } 
\left (   \frac{x(t+dt)(x,y) - x}{dt}   \right )^2 
\label{k1}
\ff
We write $x(t+dt)(x,y) $ to indicate that the position of the particle a time
$dt$ into the future is a function of both $x$ and the hidden variables $y$.  
We write the square from now on as a shorthand. 

Since the path in the joint configuration space of $x$ and $y$ is assumed to be differentiable we can write this just as well as
\begin{eqnarray}
k(x,y) & =& \frac{m}{4} \lim_{dt \rightarrow 0 } 
\left [ \left (   \frac{x(t+dt)(x,y) - x}{dt}   \right )^2 \right. \nonumber \\
&& + \left. \left (   \frac{x - x(t-dt)(x,y) }{dt}   \right )^2 
\right ]
\label{k2}
\end{eqnarray}

We will see shortly that the second form, (\ref{k2}) is preferable  because it preserves invariance under time reversal for finite $dt$.  This can matter because, once we average over the ensemble of hidden variables,  the paths in configuration space of the subsystem are Brownian and hence are no longer differentiable.

We can then write
\begin{eqnarray}
<k>  & =& \int dy \int d^N x  \tilde{\rho} (x,y) \frac{m}{4} \lim_{dt \rightarrow 0 } 
\left [ \left (   \frac{x(t+dt)(x,y) - x}{dt}   \right )^2 \right. \nonumber \\
&& + \left. \left (   \frac{x - x(t-dt)(x,y) }{dt}   \right )^2 
\right ]
\label{kinetic1}
\end{eqnarray}

We now want to integrate over the hidden variables, which requires that we express
the dynanics of the positions $x^a$ not as ordinary variables but in terms of 
stochastic differential equations.  That is, we assume that the dependence on 
the hidden variables $y$ is responsible for the variance in an ensemble
over trajectories in the $x^a$.  This lets us write.
\f
x^a(t+dt)(x,y) - x^a = b^a(x,t) dt + dw^a (x,y)
\label{trade2}
\ff
where $b^a(x,t)$ is not a function of the hidden variables but the fluctuations
$dw^a$ are.  Similarly we write 
\f
x^a - x^a(t-dt)(x,y)  = b^{*a} (x,t) dt + dw^{*a} (x,y)
\label{trade3}
\ff

We then assume we can integrate over the $y$'s , giving an ensemble of Brownian trajectories satisfying the earlier equations such that
\f
\int dx dy \rho (x,y) F(x,y) = \left \langle F \right \rangle
\ff
However, before doing the $y$ integral in eq. (\ref{kinetic1}) we have
to {\it exchange the order of that integral with the limit $dt \rightarrow 0$.}
This is because the individual paths of the Brownian particles are not differentiable. 

This exchange of orders of integration and a limit is a crucial assumption.
The result, as we will now see,  is to trade the dependence of the averaged energy 
on the time derivative of an individual path with a dependence on the spatial
derivative of the probability distribution of the ensemble of Brownian paths.  

Just when is this exchange of limits and integral allowed?  We can recall that
the notion that paths of Brownian motion particles are not differentiable is
meant to be an approximation good when averaging over time scales long
compared to some microscopic time scale $\tau$. For times smaller than
$\tau$ the paths can be resolved into differentiable paths, while for time scales
longer than $\tau$ there is so much chaos in the paths that it is most useful to understand the dynamics in terms of an ensemble of Brownian particles.

The crucial step is the change of description from one  in terms of motion on a joint configuration space of observables and hidden variables to a Brownian ensemble of
trajectories just of the $x^a$'s.  This should then be understood as a change of
veiwpoint from following the joint trajectories precisely in time and one in which time is not resolved below the scale $\tau$.

The exchange of order of limit and integral has to be understood in this light. 
By the assumption that 
\f
\int dy \lim_{dt \rightarrow 0 }  =    \lim_{dt \rightarrow 0 }  \int dy
\ff
we mean more precisely that an approximation is being made in which 
\f
\int dy \lim_{dt \rightarrow 0 } \approx   \lim_{dt \rightarrow \tau }  \int dy.  
\ff

Let us make this assumption and see  where it leads. We then have
from (\ref{kinetic1})
\begin{eqnarray}
&& <k>   =  \lim_{dt \rightarrow 0 } \frac{m}{4}  \int dy \int d^N x  \tilde{\rho} (x,y) 
 \\
 && \left [ \left (   \frac{x(t+dt)(x,y) - x}{dt}   \right )^2 
 +  \left (   \frac{x - x(t-dt)(x,y) }{dt}   \right )^2 
\right ]  \nonumber 
\label{kinetic2}
\end{eqnarray}
Using (\ref{trade2},\ref{trade3}) this is
\begin{eqnarray}
&& <k>   =  \lim_{dt \rightarrow \tau } \frac{m}{4}  \int dy \int d^N x  \tilde{\rho} (x,y) 
 \\
 && \left [  ( \frac{b^a(x,t) dt + dw^a (x,y)}{dt}   )^2 
 +   (  \frac{ b^{*a}(x,t) dt + \Delta^* w^a (x,y) }{dt}    )^2 
\right ]  \nonumber 
\label{kinetic3}
\end{eqnarray}
We can then take the average and the limit to find
\begin{eqnarray}
<K>& =&  \frac{m}{4} \int d^N x \rho (x) \left [ b^2 + b^{*2}  \right ]+ C + O(\tau)
\nonumber \\
&=& 
\frac{m}{2} \int d^N x \rho (x) \left [ v^2 + u^{*2}  \right ] + C + O(\tau )
\label{finalK}
\end{eqnarray}
where $O(\tau )$  denotes terms that vanish as $\tau \rightarrow 0$ and $C$ is a constant
\f
C= \frac{\nu m}{2 \tau}
\ff
Should we worry that the averaged conserved energy shifts by a constant that becomes
infinite as $\tau \rightarrow 0$?  There is nothing to worry about because we are 
employing the principle that the average energy is conserved in time. Shifting it by a large or even infinite constant does not affect its constancy in time. 

\subsection{The importance of time reversal invariance}

There is  a subtle issue buried in the derivation we have just described that needs to be discussed,  as it
makes clear the importance of the assumption of time reversal invariance.  Assuming that the paths are smooth
it seems we could equally well express the kinetic energy as (\ref{k1}) or
(\ref{k2}) or in general as
\begin{eqnarray}
k(x,y) & =& \frac{m}{4} \lim_{dt \rightarrow 0 } 
\left [ \alpha \left (   \frac{x(t+dt)(x,y) - x}{dt}   \right )^2 \right. \nonumber \\
&& + \beta \left.  \left (   \frac{x - x(t-dt)(x,y) }{dt}   \right )^2 
\right ]
\label{k3}
\end{eqnarray}
where $\alpha + \beta =1$.  As long as the paths are smooth these are equal
for any choice of $\alpha$.  But following the derivation through this would lead to
\f
<K> =
\frac{m}{2} \int d^N x \rho (x) \left [ v^2 + u^{*2}  +2  (\alpha-\beta ) 
v\cdot u \right ] + C
\ff
This is not equal to (\ref{finalK}).   There must then be something wrong with the argument.  The point is that, as emphasized above, at the point where the exchange of a limit and an integral is made, the expressons are no longer equalities but approximations, good only to the extent that the exchange of orders is permissible.
This is permissible only if we are studying quantities averaged in time over a time scale
greater than $\tau$, such that the exchange of the limit and the integral is a good approximation.  

The correct values of $\alpha$ and $\beta$ then depend on the physics that allows
the exchange of the limit and integral.  Here the assumption of time reversal invariance
plays the key role. This is a physical assumption, because it affects the outcome.  The point is that for finite $dt$ the expressions (\ref{k1}) is not invariant under time
reversal, whereas (\ref{k2}) is.  If we insist that the physics respects time reversal
invariance then the derivation must respect this assumption at each step and we
must choose ({\ref{k2}). This leads to the result (\ref{finalK}), which is the form of the
averaged conserved energy that gives the Schroedinger equation\footnote{In Nelson's original formulation\cite{Nelson-PhysRev} the principle  of time reversal invariance is imposed in the definition
of the stochastic acceleration.}.  

The conclusion is that quantum mechanics is a consequence of the more fundamental theory envisioned in these derivations only if the effect of the hidden variables on the observables does not disturb the time reversal invariance.  We note that in many circumstances in which a system is put into interaction with a reservoir, the effect of the reservoir is to make the system tend to equilibrium, which gives a preferred direction of time to the thermodynamics of the system.  If hidden variables are responsible for quantum dynamics, their effect must be to disorder the dynamics of the observable quantities in a way that preserves both the conservation of an average conserved energy and the time reversal invariance of the evolution of the ensemble representing the system.

\section{Answering WallstromÕs objections to Nelson}

In \cite{Wallstrom} Wallstrom made an objection to Nelson's stochastic mechanics,
which seems at first formidable, but has a straightforward 
answer, at least for simple cases.  

To discuss Wallstrom's objection it is simplest to consider the case of quantum mechanics on a circle. The independent variables of Nelson's formulation are then
a current velocity $v (\theta )$ and a probability density $\rho (\theta )$. These
are defined on a circle parameterized by $\theta$. These are connected by the law
of current conservation
\f
\dot{\rho} = \partial (\rho v)
\label{s1-cons}
\ff

 We then have locally-but not necessarily globally,
\f
v= \frac{1}{m} \partial S
\ff
for a function $S$.  The local evolution equations are, in addition to (\ref{s1-cons}),
\f
\dot{S}= \frac{1}{2m} (\partial S)^2 + {\cal U} + \frac{b \nu^2}{2}
\left ( \frac{(\partial \rho )^2}{\rho^2} + 2 \partial ( \frac{1}{\rho} \partial \rho ) 
\right )
\label{s1-Sdot}
\ff
Let us consider the simple case where ${\cal U}=0$. There are then simple solutions of the form 
\f
\rho = \frac{1}{2\pi}   \ \ \ \ \,  \ \ \ \ \  v (\theta ) = w
\ff
for any constant current velocity $w$. These correspond to a constant density and the particle moving around the circle with an arbitrary velocity.  Locally, we then have
\f
S= wm \theta + \omega t
\ff
The dynamical equation (\ref{s1-Sdot})  fixes the frequency
\f
\omega = \frac{m w^2}{2}
\ff
One then forms
\f
\psi (\theta , t) = \sqrt{\rho}e^{\imath \frac{S}{\hbar}} =  
\frac{1}{\sqrt{2\pi}} e^{ \frac{\imath }{ \hbar} ( mw\theta - \omega t )}
\label{S1-psi}
\ff

Wallstrom's objection is then that , {\it 
$\psi (\theta , t)$ is not a quantum state because it is not single valued and smooth on $S^1$.  }

We may note, as a first reply, that 
this does not apply to the ground state where $S=0$.  Even if Nelson's procedure is able to only construct the ground state that is enough to reproduce perturbative  quantum field theory from the expectation value of local observables.

However, there is a more general response to the objection. It is simply incorrect,
because  there is absolutely no reason a quantum state cannot be discontinuous at one point, or at many points. The Hilbert space is ${\cal L}^2 (S^1 )$, and it is well known that almost every state in ${\cal L}^2 (S^1 )$ is discontinuous at one or even an infinite number of points.   In fact,  the  states (\ref{S1-psi}) are normalizable elements
of ${\cal L}^2 (S^1 )$
and their current velocities are well defined everywhere on $S^1$.  

Since the states  (\ref{S1-psi}) are in the Hilbert space,  they must correspond to solutions to Nelson's equations, if Nelson's formulation is to correspond to quantum mechanics.   Thus, these states, rather then being counterexamples to the correspondence are a  necessary consequence of it. 

What is true is that the state  (\ref{S1-psi}) is not an eigenstate of momentum.  
This is because it fails to satisfy 
\f
[ \hat{p}  \Psi ] (\theta )  = [-  \imath \hbar \partial \Psi ] (\theta ) = \hbar m w \Psi (\theta )
\ff
at every point on the circle. 
Were it
an eigenstate of momentum $w$ would be quantized. But a generic state cannot satisfy this condition, all that is required is that it be equal to a superposition of eigenstates, which by Fourier's theorem  it is.

Furthermore,  $\Psi (\theta )$ has a well defined evolution under the Schroedinger equation. 
It is a solution to (\ref{s1-cons}) and (\ref{s1-Sdot}), which are just the real and imaginary parts of the 
Schroedinger equation.  

Wallstrom's objection is thus answered for the case of quantum mechanics on the circle.  The objection has been made as well, for the case of wavefunctions on a general configuration space. Bacciagaluppi and  Valentini have suggested that the map between solutions to Nelson's equations and states in the quantum Hilbert space could  have 
obstructions or ambiguities in cases where the quantum state has several 
nodes \cite{antony-personal}, but this issue has not been treated in detail.  One may also argue that states must be differentiable for the action of the Hamiltonian to be defined, which would imply that most of ${\cal L}^2$ of a configuration space are not good quantum states.  However, as  the above example illustrates,  both the expectation value of the energy and the time evolution are well defined for all solutions to Nelson's equations. Hence, Nelson's formulation may be argued to provide a definition of the action of the Hamiltonian for such cases.

\section{Conclusions}

In this paper we have made a careful study of the question of when quantum theory can be derived from a non-local hidden variables theory, as an approximation to another, non-quantum theory.  This is relevant for attempts to derive quantum theory from non-local hidden variables theories as in 
\cite{hidden,morehidden,Adler}.    While the discussion hinged on some seemingly technical points, it brought out key conditions that must be satisfied if such a derivation is to succeed.

The scene of the crime, so to speak, is the origin of the term which leads to the quantum potential,
(\ref{hquantum}).  This is the crucial term in Bohm's formulation by which the wavefunction guides the particle, it is hence the reason why the wavefunction has, in Bohm's formulation, to be considered as real as the particle. We see that in Nelson it gives the probability density a non-classical role as it gives a contribution to the energy of an individual system that depends on the gradient of its probability density.  Hence, by the presence of this term, Nelson's formulation seems to give the ensemble a non-classical meaning, thus failing to avoid the doubled ontology of Bohm.

Could this issue be resolved by deriving the term from a non-local hidden variables theory? We gave a derivation suggesting that it could be. However a look at the details  reveals the conditions that must be satisfied for the derivation to succeed. Among these  is that the stochastic dynamics of the subsystem which results have strict averaged energy conservation and invariance under time reversal.    This is not impossible, but we note that it is contrary to  the expectations suggested by our experience with  ordinary statistical mechanics. There, coupling to a reservoir of degrees of freedom results in violations of both energy conservation and time reversal invariance.  

However in the approaches studied in \cite{Adler,hidden}, we are in a very different situation from ordinary thermodynamics, because the interactions between the reservoir and system are mediated, not by heat flow across a wall,  but by a network of non-local interactions. These are given  by off diagonal matrix elements in Adler's formulation and by explicit non-local interactions in the formulation in \cite{hidden}.  It is then not impossible that the conditions found here to be necessary could be satisfied\footnote{It should be stressed that while it is natural to mention Adler in this context, the exact
relation between Adler's formulation and stochastic mechanics has not been, to my knowledge, established.  Thus, there as yet no claim that the conclusions of this
paper apply to Adler's formulation.}.

Finally, we discussed the objection of Wallstrom\cite{Wallstrom} which is often mentioned as a reason why Nelson's formulation must be inequivalent to quantum theory.  We see by studying the example of quantum mechanics on a circle that the argument fails in that case. Instead, the solutions of Nelson's formulation thought to be problematic are required because they correspond to states whose wavefunctions are discontinuous but, being in ${\cal L}^2 (S^1)$ are  in the Hilbert space of quantum theory.

\begin{acknowledgements}

Many thanks to Guido  Bacciagaluppi,  Joy Christian, Olaf Dreyer,  Shelly Goldstein, Lucien Hardy, Fotini Markopoulou, Simon Saunders, Ward Struyve and  Antony Valentini for lively discussions on these issues. I am also grateful to Sabine Hossenfeldere,  Ward Struyve and Antony Valentini for comments on the manuscript and to Steven Adler for conversations on his approach.  The work of section IV was initially begun with Poya  Haghnegahdar, to whom I owe many thanks for his efforts towards clarifying these issues.   Research at Perimeter Institute for Theoretical Physics is supported in
part by the Government of Canada through NSERC and by the Province of
Ontario through MEDT.
 
 \end{acknowledgements}

\end{document}